\documentclass[a4paper]{article}

\usepackage{amsmath}

\usepackage{oljour}
\usepackage{graphicx}
\usepackage{hyperref}

\usepackage{epsfig}
\usepackage[english]{babel}

\begin{document}

\journalname{it}  


\title[en]{Can electoral popularity be predicted using socially generated big data?}

\begin{author}
  \anumber{1}   
  \atitle{Dr.}    
  \firstname{Taha}  
  \surname{Yasseri}    
  \vita{Taha Yasseri holds a PhD from the institute of 
  theoretical physics at the University of G\"ottingen, Germany, where he worked on spontaneous pattern formation in 
  complex systems followed by two years of Postdoctoral Research at the Budapest University of Technology and Economics, 
  working on the socio-physical aspects of the community of Wikipedia editors. He is now a researcher at 
  the Oxford Internet Institute, University of Oxford. His main research focus now is on online societies, government-citizen 
  interactions on the web and structural evolution of the web. He uses mathematical models and data analysis to study social 
  systems quantitatively. }      
  \institute{Oxford Internet Institute, University of Oxford}
  \street{1 St Giles'}
  \number{}
  \zip{OX1 3JS}
  \town{Oxford}
  \country{UK}
  \tel{+44-1865-287229}      
  \fax{+44-1865-287211}      
  \email{taha.yasseri@oii.ox.ac.uk}
\end{author}

\begin{author}
  \anumber{2}   
  \atitle{Dr.}    
  \firstname{Jonathan}  
  \surname{Bright}    
  \vita{Jonathan Bright is a political scientist specialising in computational and ``big data'' approaches to the social sciences. 
  He holds a BSc in Computer Science from the University of Bristol, an MSc in International Politics from the School of Oriental 
  and African Studies, and a PhD in Political Science from the European University Institute.
His major research interests lie in both the quantitative study of online news media (and, more generally, virtual ``public spheres''), 
and also the large scale analysis of politicians and parliamentary behaviour. More generally, he is also interested in developing 
computational social research methods, and in driving forward computing skills as a core part of social science methodology.}      
  \institute{Oxford Internet Institute, University of Oxford}
  \street{1 St Giles'}
  \number{}
  \zip{OX1 3JS}
  \town{Oxford}
  \country{UK}
  \tel{+44-1865-287233}      
  \fax{+44-1865-287211}      
  \email{jonathan.bright@oii.ox.ac.uk}
\end{author}

\corresponding{taha.yasseri@oii.ox.ac.uk}

\abstract{Today, our more-than-ever digital lives leave significant footprints in cyberspace. 
Large scale collections of these socially generated footprints, often known as big data, could 
help us to re-investigate different aspects of our social collective behaviour in a quantitative 
framework. In this contribution we discuss one such possibility: the monitoring and predicting of 
popularity dynamics of candidates and parties through the analysis of socially generated data on the web during electoral campaigns.
Such data offer considerable possibility for improving our awareness of popularity dynamics. However they also suffer from significant 
drawbacks in terms of representativeness and generalisability. In this paper we discuss potential ways around such problems, suggesting 
the nature of different political systems and contexts might lend differing levels of predictive power to certain types of data source. 
We offer an initial exploratory test of these ideas, focussing on two data streams, Wikipedia page views and Google search queries. On 
the basis of this data, we present popularity dynamics from real case examples of recent elections in three different countries. }


\keywords{E.m [ Data:Miscellaneous]; J.4.c [Applications: Social and Behavioral Sciences: Sociology]; J.8.o [Applications: Internet Applications:Traffic analysis]; J.2.h [Applications: Physical Sciences and Engineering: Mathematics and statistics]; J.8.j [Applications: 
Internet Applications: Libraries/information repositories/publishing]; K.4.3.b [Computing Milieux: Computers and Society: Computer-supported collaborative work]}

\received{} \accepted{} \volume{} \issue{} \class{} \Year{}

\maketitle


\section{Introduction}
Increasing use of the internet, and especially the rise of social media, has generated vast quantities of data on human behaviour,
significant portions of which are also readily available to researchers. The potential of these data has not gone unnoticed: in just 
a few years use of social media data in particular has started to see a wide variety of applications in the growing subfield of ``computational social science'' \cite{Lazer2009, Conte2012}.
One of the most intriguing possibilities raised by the emergence of social media data is that it could be used to supplement (or even
eventually replace) traditional methods for public opinion polling, especially the sample survey, because social media data offer 
considerable advantages in comparison with surveys in terms of the speed with which they can be acquired and the cost of collection. 
The selection bias in social media is clear: not everyone uses it, and people who do are not randomly distributed throughout the 
population \cite{Mislove2011}. Yet the hope has frequently been expressed that the sheer quantity of social media users may start to compensate for 
this (around 50\% of the UK's population are thought to have a Facebook account, for instance) and hence that we might eventually replace 
the ``sample-based surveys'' with the ``whole population data''. 

The potential applications of ``social polls'' are wide ranging, however probably the most frequently explored avenue of 
research has been the use of social media data for electoral prediction. 
This is because the outcomes of elections are interesting in and of themselves, but also because it is a subject where a huge 
amount of validation data exists, coming from both the more traditional opinion polling which social media data might hope to 
replace, and the results of the election itself. Such social polling, which has largely been applied to data coming from Twitter, is typically based on one of two main methodologies: either 
offering some type of count of all tweets mentioning a given candidate (perhaps controlling for the candidate's own social media account); or using various techniques developed for analysing the 
sentiment expressed in them as a measure of people's opinion on a given candidate (see e.g. \cite{OConnor2010,Tumasjan10,Jungherr2013,Ceron2013}).

Despite initial enthusiasm, and in contrast to the cases of predicting arrival of earthquake
waves \cite{Sakaki2010} and traffic jams \cite{Okazaki2011}, most of the recent research on using Twitter for electoral prediction has been relatively negative, 
with many researchers reporting weak correlations with actual electoral outcomes, difficulty duplicating other positive research, 
or rates of successful prediction that could easily have come about by chance (see inter alia \cite{GayoAvello2011,GayoAvello2012}). 
Most problematically, results have often exhibited specific biases either for or against individual political 
candidates, with minority parties often systematically overstated (see \cite{Jungherr2013}), 
whilst major conservative candidates are often undertstated \cite{GayoAvello2012}. 

A variety of potential reasons have been put forward for these problems. The most obvious is that the self-selection 
problem of social media cannot in fact easily be overcome with a larger sample size. Self-selection also operates when
users decide what to post: even if large amounts of the population have created social media accounts, the amount which 
use them to express political opinions is much more limited. The nature of social media also means that opinions which 
are expressed are those heard by friends, family, work colleagues and other social connections: which might compel people 
to moderate their opinions or keep quiet if they support particular types of political party. Furthermore, many researchers 
have observed the difficulty of reliable sentiment analysis of political tweets, both because of the small amount of 
information contained in any given tweet and because of the nuances of political language where many opinions are expressed 
through irony or sarcasm \cite{GayoAvello2012}. Finally, as social media have started to take on a prominent position in media landscape 
(with trending topics now frequently a basis for news stories), political candidates have also increasingly started to
intervene actively in social media, which has the potential for biasing results \cite{Metaxas2012}.

\subsection*{Google Trends and Wikipedia Page Views: Predicting the Present}
While social media data are probably the most used of the new data sources which have been generated by the internet, 
significant interest has also arisen surrounding the use of informational search data present in websites such as Google 
Trends or Wikipedia, which is generated when someone either conducts a web search for a particular topic or accesses a 
particular page on Wikipedia. While not typically regarded as social ``media'', search data is nevertheless socially 
generated in that it relies on people entering individual search queries. Having clear information on what people are 
looking for and when they are looking for it provides a number of opportunities to ``predict the present'': 
to gain a kind of real time awareness of current behaviour patterns. Such data have already been used to successfully 
predict a wide variety of phenomena both in short and long terms, from car and house prices to trends in flu outbreaks or 
unemployment \cite{ Choi2009a,Choi2009b,Goel2010,Cook2011} using web search data, as well as movie box office revenues
using Wikipedia page view statistics \cite{Mestyan13}.

Information seeking data offers significant theoretical advantages when compared to social media data in terms of its use for 
prediction. Whereas the automatic interpretation of the meaning of a tweet can be riddled with complexity, the interpretation
of the meaning of a search or the access of a page in Wikipedia is much more straightforward: the user is interested in 
information on the topic in question. Furthermore, the penetration of search especially is far greater than many social 
media platforms, especially Twitter. Approximately 60\% of internet users use search engines \cite{Dutton2011}. 
For many users, Wikipedia is the most common source of knowledge online: 29.6\% of academics prefer Wikipedia to online library catalogues \cite{weller2010}, and
52\% of students are frequent Wikipedia users, even if their instructor advises them not to use the platform \cite{head2010}. In
general, browsing Wikipedia is the third most popular online activity, after watching YouTube
videos and engaging into social networking: it attracts 62\% of Internet users under 30 \cite{zickuhr2011}. The popularity of Wikipedia is also closely related to the significant importance given to it by search engines like Google; in 96\% of cases Wikipedia ranks within the top 5 UK Google search
results \cite{silverwood2012}.

However, such data has rarely been applied to the task of election prediction. The main reason for this is simple:
``queries are not amenable to sentiment analysis'' \cite{Gayo-Avello2013}. When entering a search query people express
what they are looking for, but not their opinion about the subject: indeed, given they are searching for information, 
it seems reasonable to assume that this opinion is not fully formed. Despite this problem, in this paper we argue that 
there is significant ``sentiment'' data implied in information seeking behaviour. In particular, we expect that searches for political 
candidates around election time imply that people may be considering voting for them (though these searches are also likely stimulated by 
the reception of other bits of information, especially from the mass media). This assumption is inspired by previous work connecting 
information seeking to eventual real world outcomes: for example, connecting it to eventual movie box office revenues \cite{Mestyan13}.

The problem for the purposes of prediction is that the relationship between search traffic and actual outcomes is unlikely to 
be straightforward. In fact, one of the few studies that has attempted to apply information 
seeking data to elections \cite{Lui2011} found that simply using search volume in the days prior to the election is an extremely poor prediction technique. 
Rather, we argue, there are a number of intervening variables which may affect how people look for information on politics, and 
thus need to be taken into account. One obvious first factor would be whether the political system encourages focus on parties or 
individuals (which may itself emerge through different modes of democratic organisation, e.g. presidentialism vs. parliamentarism), 
something which is likely to affect the search terms people enter. Also worth considering is the amount of potential candidates on 
the political scene, with elections full of new faces likely to generate more searching than contests between familiar 
candidates. Finally, there is the extent to which the existing incumbent is popular: as people are more likely to be informed on 
what the current power holder's views are, they are less likely to search for them. 

Within the context of this paper, we seek to explore some of these questions by looking at correlations between search engine data, 
Wikipedia usage patterns and recent election results in three different countries: the UK, Germany and Iran. These countries were 
selected in order to provide a diverse range of political contexts (with elections in Iran and the UK where a new candidate was
voted in and one in Germany where a popular incumbent was returned), electoral systems (from Iran's presidential system to the 
parliamentary ones operated in the UK and Germany) and party landscapes (with a very stable system in the UK contrasted to Iran 
and Germany where new actors are emerging).

\section{Data Collection}

For our analysis, we collected data from both Google Trends and Wikipedia for the last election in each of our countries of interest 
(2013 in the case of Iran and Germany, 2010 in the UK). Our trends data is based on the amount of searches for either a given party or 
politician coming from our specific country of interest (search terms were entered in the native language and script of that country). 

Our Google data was collected directly from the Google Trends website (\url{http://www.google.com/trends/}). This site allows users to compare the relative search volumes of different keywords, and to download the resulting data in CSV format. The specific keywords used are reported in tables 1-3. We assume that these data are reliable, as they come out of Google's own server logs.

Our Wikipedia data is extracted from the page view statistics section of the Wikimedia Downloads site (\url{http://dumps.wikimedia.org/other/pagecounts-raw}) 
through the web-based interface of ``Wikipedia article traffic statistics'' (\url{http://stats.grok.se}); again, for Wikipedia we 
focus on language specific terms appropriate to the country of interest. Although the original data dumps are of hourly granularity, in this research we
used a daily accumulation of data in GMT. While the actual logs count the url requests, they might not well represent the unique visits nor unique visitors
to the page. On the positive side, if the title of the page has been searched in alternative forms, and the user has been redirected to the page, this should have been
counted in the data. In the case of Google search volume it is more problematic, because there is no systematic way to aggregate the data for different search keywords.
For the sake of simplicity, in this work we only considered a most common keyword for each item, being aware of the biases that it might
introduce in the data. 

\section{Results}

\begin{figure}
\includegraphics[height= 6 cm]{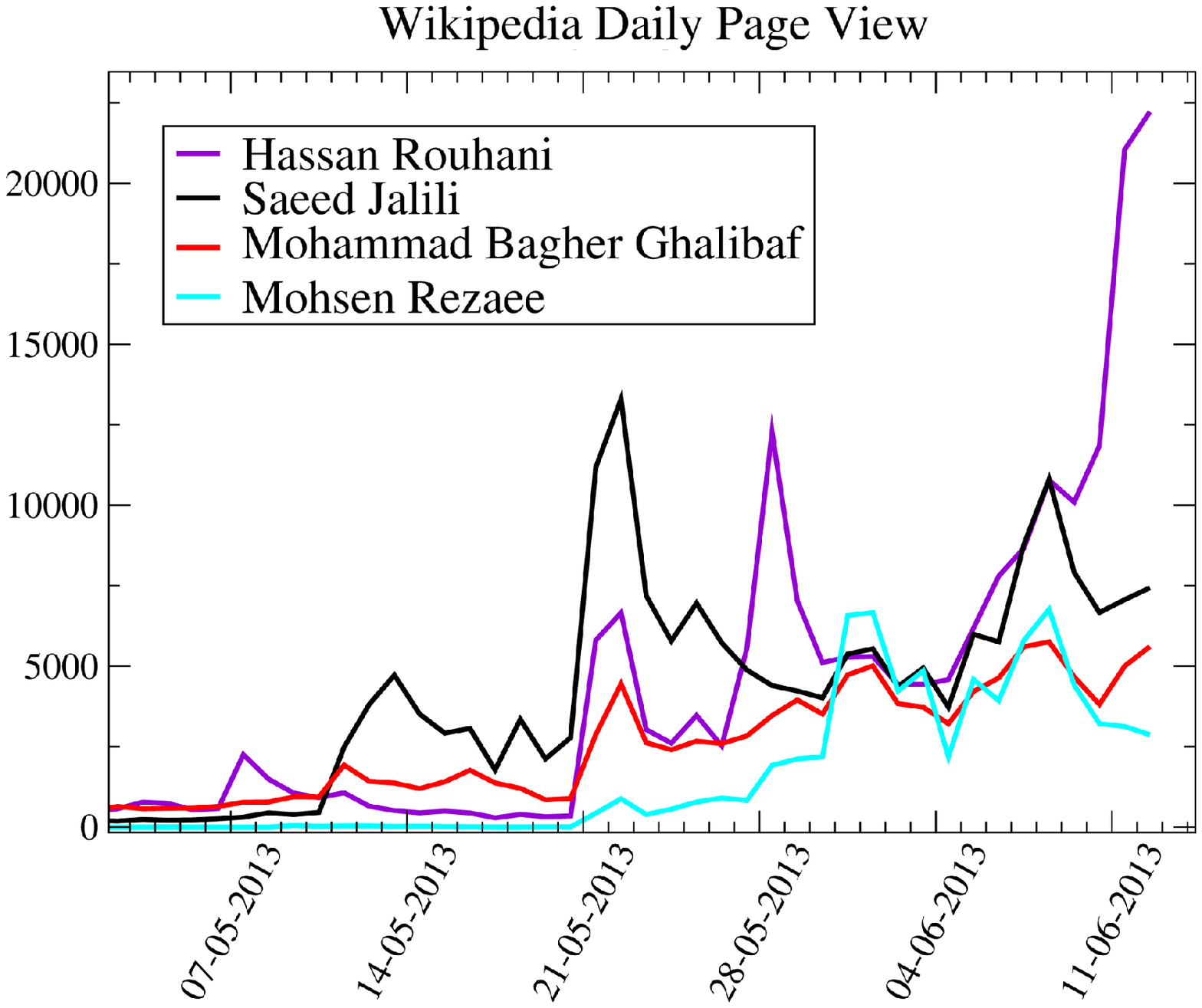} \hspace{.5 cm} \includegraphics[height= 6 cm]{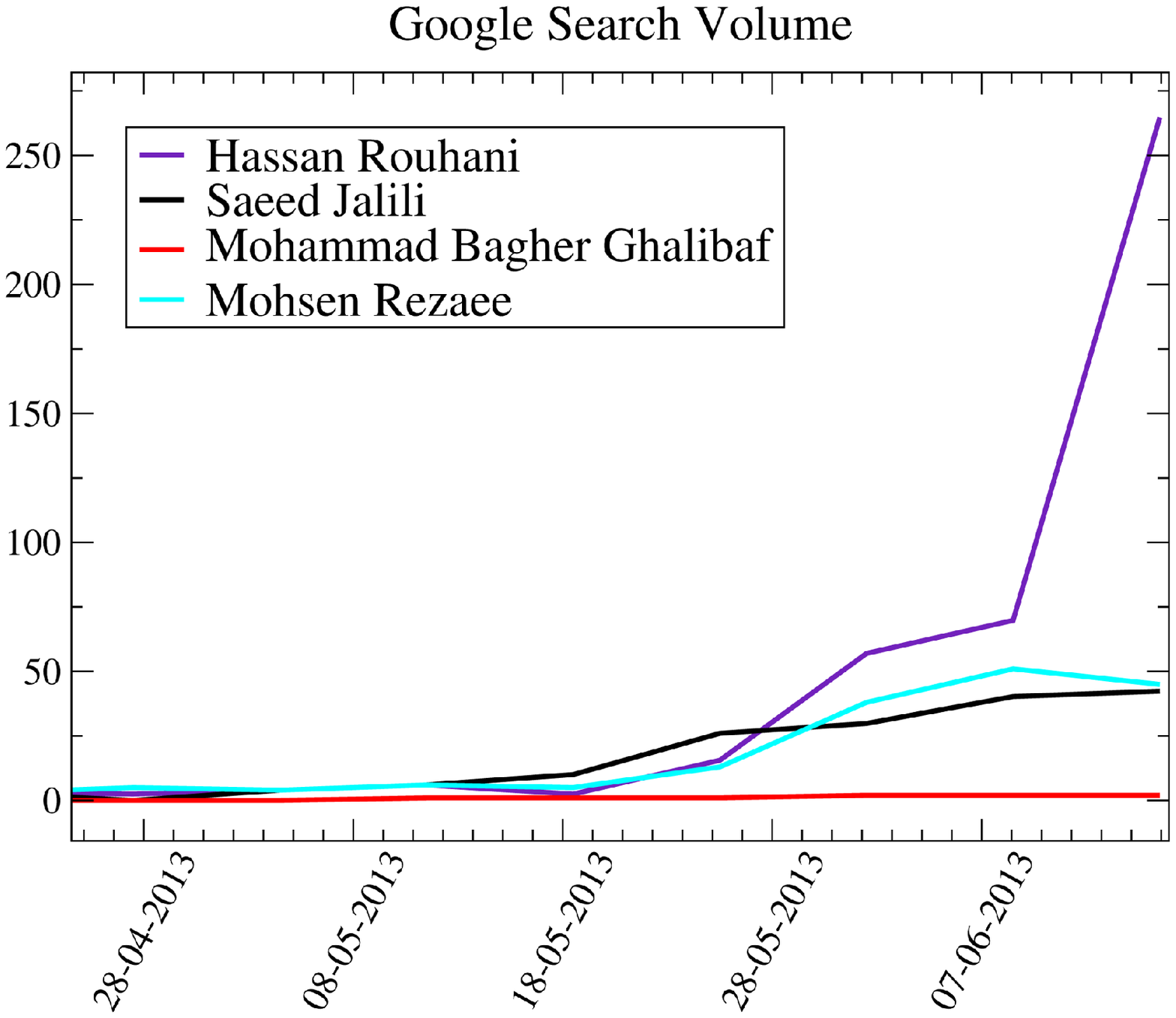}
\caption{The time evolution of Wikipedia page views and Google search volume for the four leading candidates of the Iranian presidential election of 14 June 2013
are shown in the left and right diagrams respectively.}\label{fig:1}
\end{figure}

We will begin with a discussion of the Iranian election of the 14th of June 2013. Iran operates a presidential system, where 
individual candidates are far more important than political parties. The presidency goes to the candidate who gains more than 
50\% of the vote, with a run-off in case no candidate is able to in the first round. The election of 2013 was an unusual one: 
it lacked an incumbent candidate (with former president Mahmoud Ahmadinejad standing down after fulfilling the maximum two terms 
in office), and was won convincingly by Hassan Rouhani in the first round, a candidate who was perceived as an outsider just a month before the election. 
Figure~\ref{fig:1} shows patterns in Wikipedia page views and Google Search volume for the Iranian election, whilst the final results 
can be seen in Table~\ref{tab:1}. Several patterns are immediately apparent. First, both the quantity of searches on Google and the number of page views on Wikipedia indicate the winner of 
the election correctly, and also pick up on the large absolute disparity between Rouhani and the other candidates. 
They are both also sensitive to the very late development of Rouhani as a candidate (though Wikipedia also shows a spike in May). This comes 
as a very interesting result as none of the official polls have predicted the victory of any candidate in the first round of the election (the most optimistic poll has predicted 42\% of votes
for Rouhani) \cite{bbc}. 
However, neither Google nor Wikipedia correctly identify second place. 

{\bf $>>$ Figure 1 to be placed here $<<$}

\begin{table*}
\begin{center}
\begin{tabular}{|p{4.1cm}|l|l|p{4.0cm}|p{4.0cm}|}
\hline
Candidate  & Popular Vote & Percentage&Wikipedia page title&Google search keyword\\
\hline
\hline
Hassan Rouhani  &18,613,329 & 50.88&{\it Hassan Rouhani}& ``{\it hassan rouhani}''\\
\hline
Mohammad Bagher Ghalibaf  & 6,077,292 & 16.46 &{\it Mohammad Bagher Ghalibaf}&``{\it mohammad bagher ghalibaf}''\\
\hline
Saeed Jalili  & 4,168,946 & 11.31 &{\it Saeed Jalili}&``{\it saeed jalili}''\\
\hline
Mohsen Rezaee & 3,884,412 & 10.55 &{\it Mohsen Rezaee}&``{\it mohsen rezaee}''\\
\hline
\end{tabular}
\caption{Main candidates of the Iranian presidential election, 14 June 2013. Both Wikipedia and Google data were taken from Persian Wikipedia and therefore used titles in Persian script; English translations are shown in {\it italic}.}
\label{tab:1}
\end{center}
\end{table*}

We will now move on to the German election of the 22nd of September 2013. Germany operates as a federal parliamentary republic, 
with power divided between the German parliament (``Bundestag'') and the body which represents Germany's regions (``Bundesrat''). 
This election in particular was for the Bundestag, which itself has responsibility for electing Germany's Chancellor, 
its most powerful political office. Germany's system is based strongly around parties: a majority vote is required to elect the 
Chancellor, which is usually based on a coalition between two or more parties. In this particular election, the winning Christian 
Democrat party (CDU/CSU) increased its vote share for the second successive election, confirming its place as a highly popular 
incumbent party. However its coalition partner from the 2009 elections (the FDP) lost a lot of ground, failing to win any seats, resulting eventually in the formation of a ``grand coalition'' between CDU/CSU and the major social democratic party (SPD). 

\begin{figure}
\includegraphics[height= 6 cm]{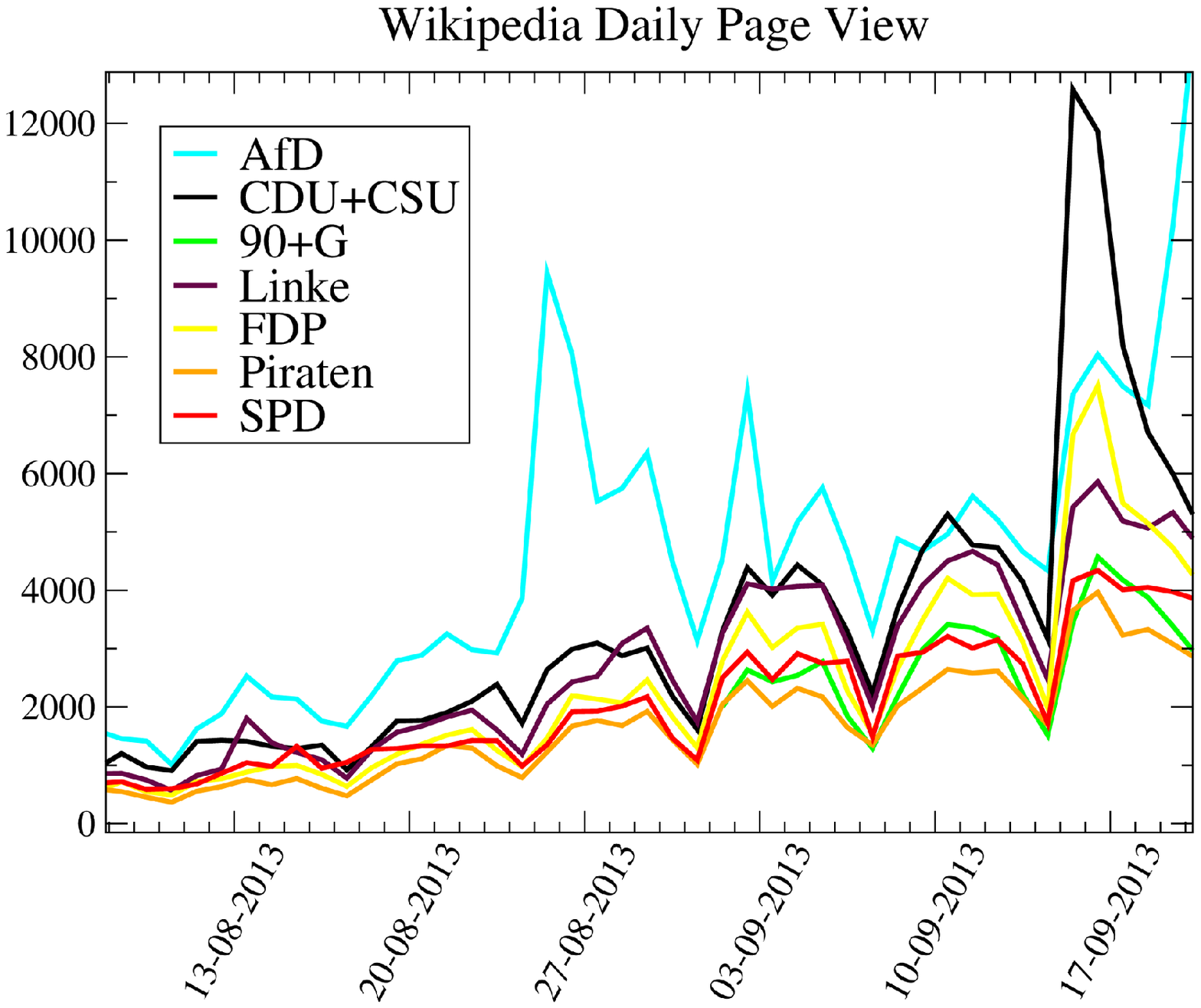} \hspace{.5 cm} \includegraphics[height= 6 cm]{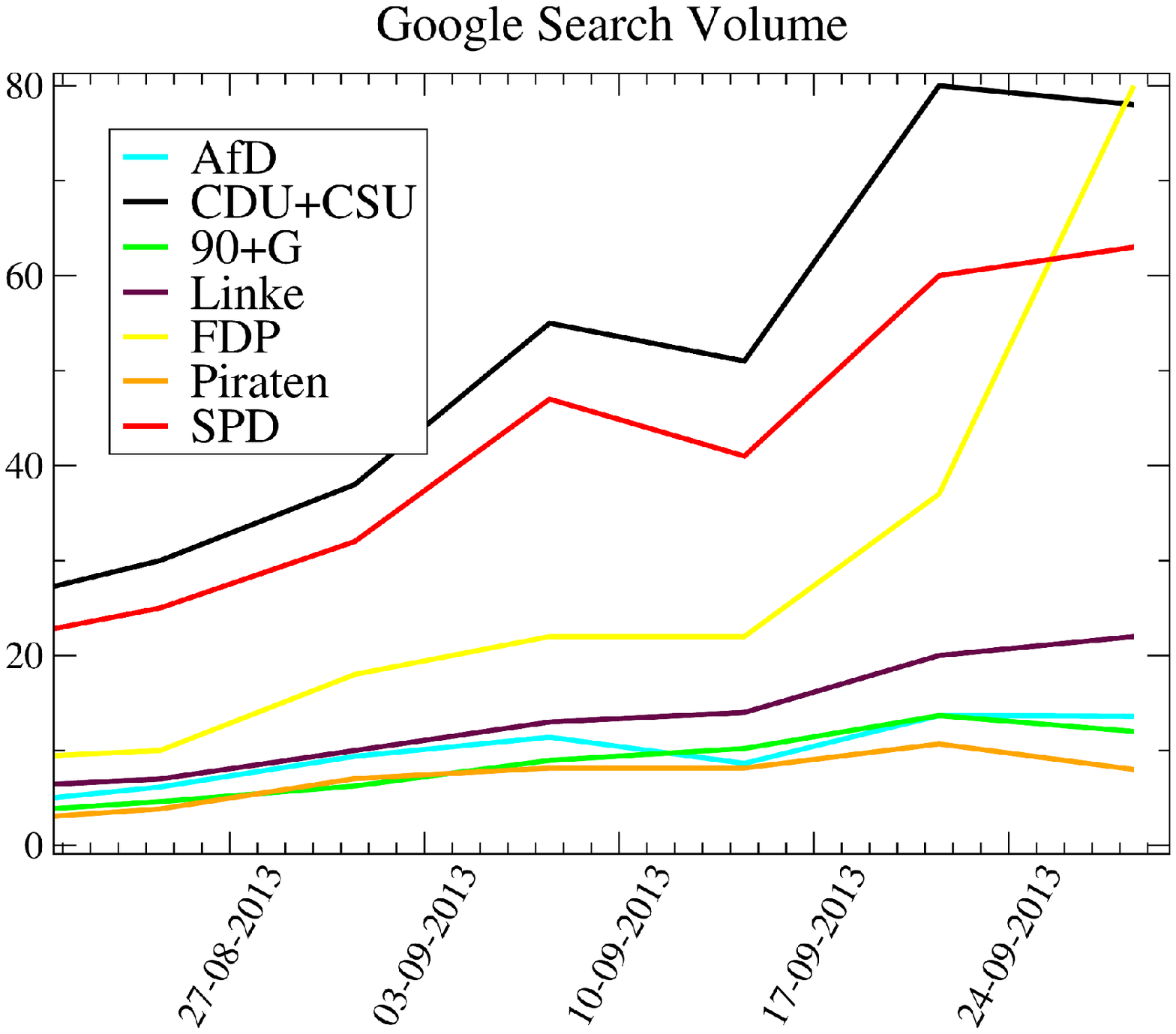}
\caption{The time evolution of Wikipedia page views and Google search volume for the 7 leading German parties during the 22 September 2013 parliamentary election campaign are shown in the left and right diagrams respectively. Note that the figures for CDU + CSU and Alliance 90 + The Greens (90+G) are produced by summing Wikipedia page views and Google Searches for each party name individually}\label{fig:2}
\end{figure}

The results of the election are shown in Table~\ref{tab:2}, whilst the data extracted from Wikipedia and Google are shown in Figure~\ref{fig:2}. 
The results show an interesting contrast to the Iranian election. Google predicts correctly both the winner of the election and second 
place (if we look at the date of the election), and is also approximately right about the distance between the two parties. 
It radically overstates the position of the FDP however. Wikipedia, by contrast, does not predict anything accurately, overstating 
to a large extent the position of Alternative for Germany (AfD), a radical anti-Euro party which was recently formed. This chimes 
with earlier work by Jungherr \cite{Jungherr2013} who found that Twitter overstated to a large extent the position of the Pirate Party (which was also recently formed) in the 2009 German election. 

{\bf $>>$ Figure 2 to be placed here $<<$}

\begin{table*}
\begin{center}
\begin{tabular}{|p{4.1cm}|l|l|p{4.0cm}|p{4.2cm}|}
\hline
Party & Popular Vote & Percentage&Wikipedia page title&Google search keyword\\
\hline
\hline
Christian Democratic Union & 14,921,877 & 34.1&Christlich Demokratische Union Deutschlands&cdu \\
\hline
Social Democratic Party & 11,252,215 & 25.7 &Sozialdemokratische Partei Deutschlands&spd\\
\hline
The Left & 3,755,699 & 8.6 &Die Linke&``die linke''\\
\hline
Alliance '90/The Greens & 3,694,057 & 8.4&B\"undnis 90 ; Die~Gr\"unen&``b\"undnis~90''; ``die~gr\"unen''\\
\hline
Christian Social Union of Bavaria  & \,243,569 & 7.4 &Christlich-Soziale Union in Bayern&csu\\
\hline
Free Democratic Party  & 2,083,533 & 4.8 &Freie Demokratische Partei&fdp \\
\hline
Alternative for Germany & 2,056,985 & 4.7 &Alternative f\"ur Deutschland&``alternative f\"ur deutschland''\\
\hline
Pirate Party  & 959,177 & 2.2&Piratenpartei Deutschland&piratenpartei\\
\hline
\end{tabular}
\caption{Main parties of the German federal election, 22 September 2013. Note that for Alliance ’90/The Greens two separate Wikipedia pages and Google Search terms were used, which are then summed together in the analysis below.}
\label{tab:2}
\end{center}
\end{table*}

We will now look finally at the results of the 2010 UK election. The UK also operates a parliamentary system, though unlike Germany does not 
have a separate regional body. Rather, power is concentrated on one legislative body (the House of Commons), with a secondary unelected body 
(the House of Lords) providing some checks and balances. The history of the UK has been dominated by single party government, as the voting 
system there favours the emergence of a small group of very large parties. Hence even though in theory parliament and hence parties elect 
the prime minister, in practice the individual personalities of leaders have come to be seen as just important as party identity. For this 
reason in the UK we look at both individuals and parties. 

\begin{figure}
\includegraphics[height= 6 cm]{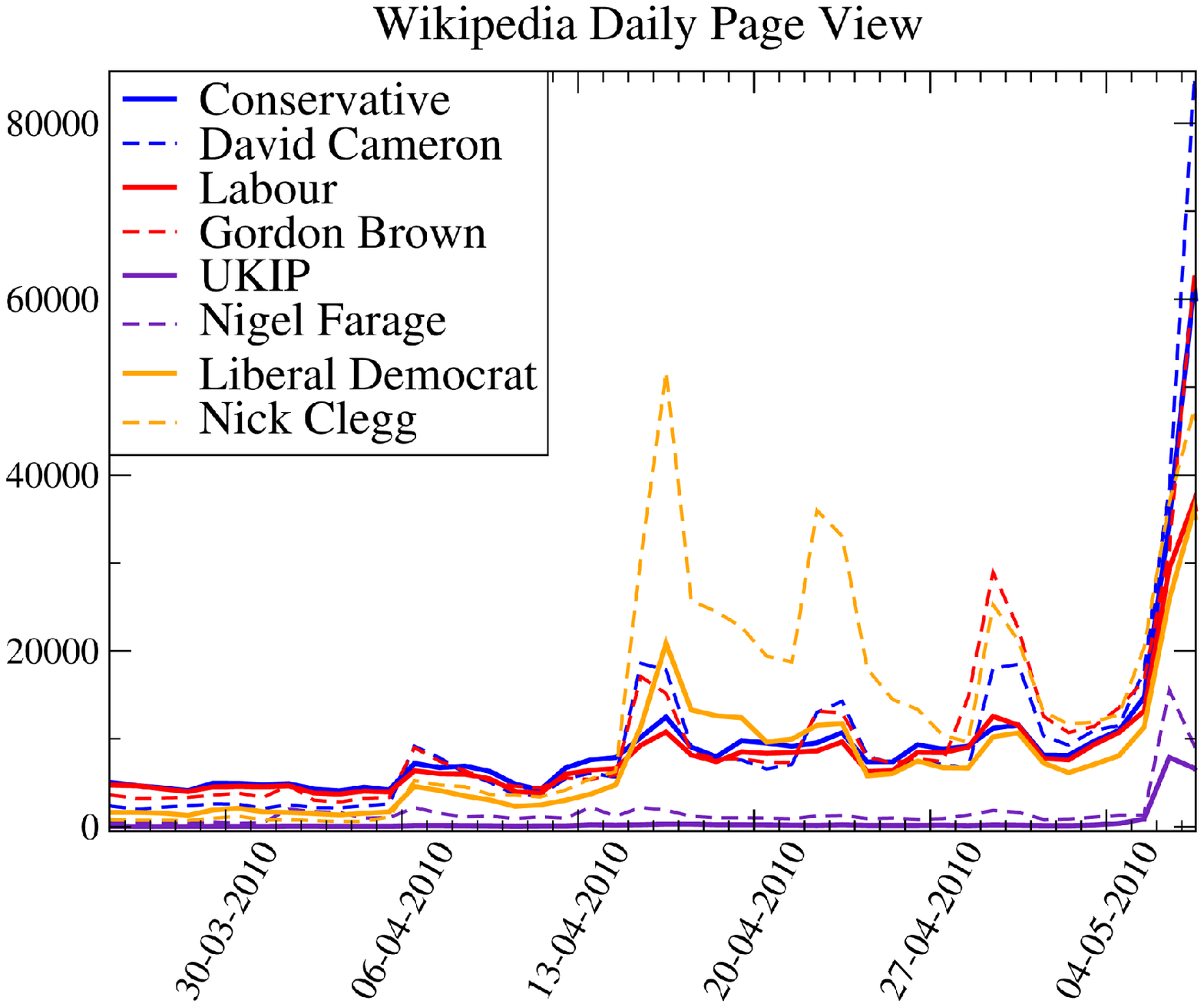} \hspace{.5 cm} \includegraphics[height= 6 cm]{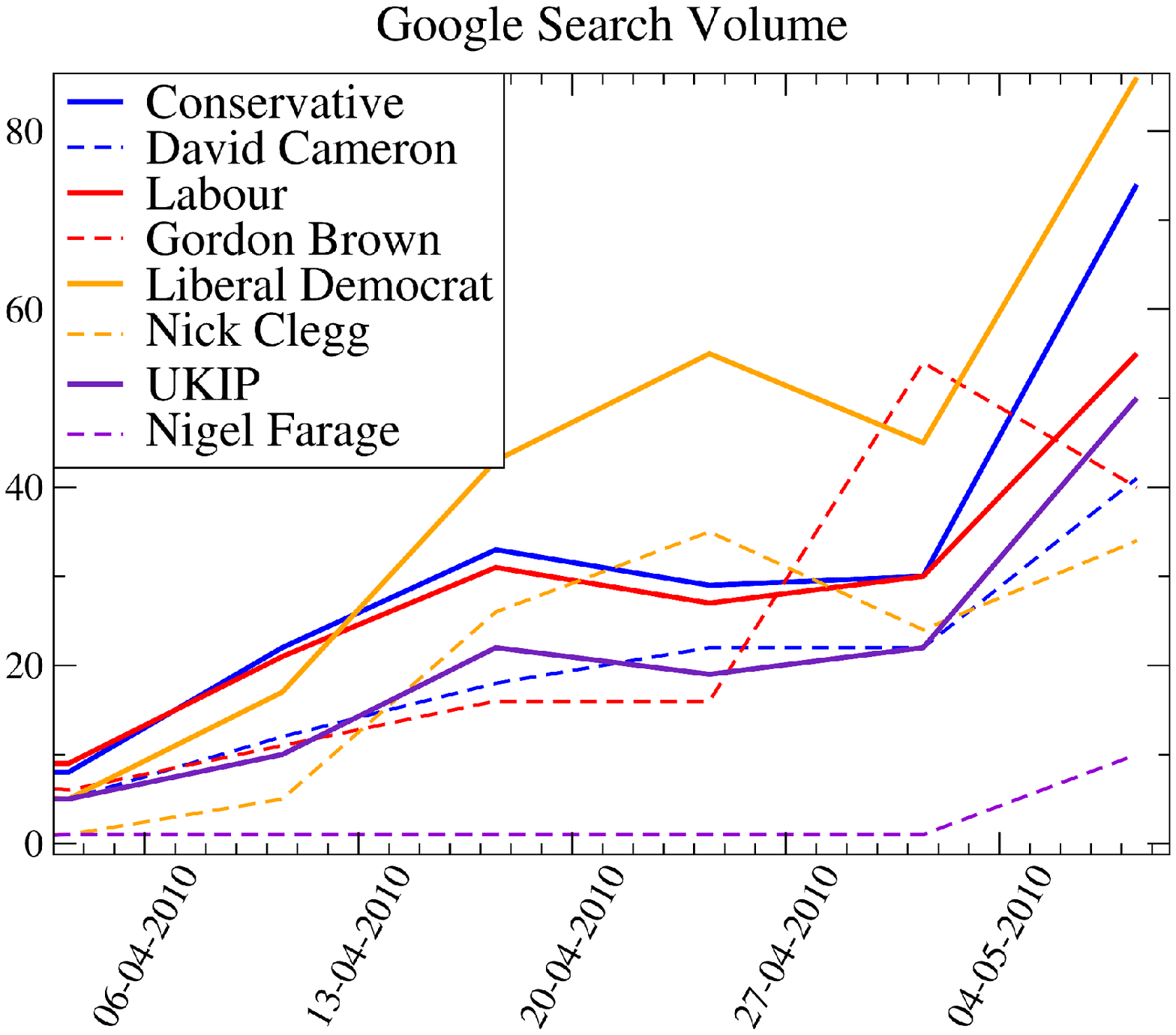}
\caption{ The time evolution of Wikipedia page views and Google search volume for the four leading parties and their leaders during the 6 May 2010 UK general election campaign are shown in the left and right diagrams respectively.}\label{fig:3}
\end{figure}

Figure~\ref{fig:3} shows results from Wikipedia and Google for the UK election, whilst Table~\ref{tab:3} reports the actual results. A variety 
of findings are worth noting here. Firstly, on Google, parties were universally more searched for than politicians, however the party 
data itself did not offer a useful predictor of the election results, considerably overstating the position of the Liberal Democrats, 
the UK's third largest party (though this party did improve considerably on its 2005 result). The individual politician data did, by 
contrast, place all the winning parties in correct order, though the difference between Conservative candidate David Cameron and Labour 
candidate Gordon Brown was marginal. In Wikipedia, by contrast, individual politicians were much more viewed than parties. Both the 
politician and party data offers a correct placement of all four parties, though the differences between them are microscopic.

{\bf $>>$ Figure 3 to be placed here $<<$}

\begin{table*}
\begin{center}
\begin{tabular}{|p{4.1cm}|l|l|p{4.0cm}|p{4.0cm}|}
\hline
Party/Leader & Popular Vote & Percentage&Wikipedia page title&Google search keyword\\
\hline
\hline
Conservative  & 10,703,654 & 36.1&Conservative Party (UK)&``conservative party''\\
\hline
David Cameron&&&David Cameron&``david cameron''\\
\hline
Labour & 8,606,517 & 29.0 &Labour Party (UK)&``labour party''\\
\hline
Gordon Brown&&&Gordon Brown&``gordon brown''\\
\hline
Liberal Democrat  & 6,836,248 & 23.0&Liberal Democrats&``liberal democrats''\\
\hline
Nick Clegg&&&Nick Clegg&``nick clegg''\\
\hline
UKIP  & 919,471 & 3.1&UK Independence Party&ukip\\
\hline
Nigel Farage&&&Nigel Farage&``nigel farage''\\
\hline

\end{tabular}
\caption{Main parties and party leaders of the United Kingdom general election, 6 May 2010.}
\label{tab:3}
\end{center}
\end{table*}

\section{Discussion and Conclusion}

There are several broad conclusions we would like to draw from this data. It is clear first and foremost that online information seeking forms a 
part of contemporary elections: all three of the countries under study showed significant increases in traffic in the days leading up to an election. 
However it is also clear that patterns differ in the context of different elections, and that people do not simply search in the same proportions 
that they vote. Even the overall patterns show dissimilarities, while German data shows a clear weekly pattern, with the minimum of volumes during weekends, 
such patterns are absent in other two countries. 

We highlight several key factors here. Firstly, data based on individual politicians proved more reliable than data based on parties: 
both Wikipedia and Google predicted the winners of the Iranian and UK elections when using individual politicians as search terms. This may be 
because there is a greater variety of ways in which people can search for information on a political party than there is on an individual 
(they could, for example, use an abbreviation, or search for ``Labour Party'' rather than ``Labour''). However it is also interesting to note that the absolute volume of searches for parties was higher than it was for candidates in the UK case. Overall, this may mean that predictions based on social data may perform better in political systems which encourage a focus on individuals. Further research would be needed to establish these reasons more systematically.  

Secondly, it is clear that information seeking data reacts quickly to the emergence of new ``insurgent'' candidates, such as Hassan Rouhani or the AfD. 
However, supporting previous work, it may also overstate them (the high volumes for the Liberal Democrats in the UK can also be read in this light). 
For this reason, it may be useful for social predictions to look for multiple different information sources. The AfD, for example, performed well in 
Wikipedia but poorly on Google, whilst the reverse was true for the Liberal Democrats. Rouhani, by contrast, performed well on both platforms. This indicates as well that Google and Wikipedia are put to slightly different uses: the high level of AfD searches on Wikipedia suggesting that it is a key resource for people who are unaware of the views of new political forces.

Finally, it seems that information seeking data is at its least effective when predicting the decline of a previously popular party. The FDP provides 
the example here: there is little to suggest in either Google or Wikipedia that it was about to suffer the reverse it did. It may be that, as the decline of the party itself becomes newsworthy, people increase their information seeking activity on the party to find out more about why people aren't supporting it; though again further research would be required to establish this.

In conclusion, we argue that there is significant potential in information seeking data for both enhancing our knowledge of how contemporary 
politics work and predicting the outcome of future elections. It also has considerable potential benefits in comparison with social media data, as it requires no complex sentiment detection. However much work remains to be done in establishing the conditions under which 
such prediction will be successful. In our view, this will depend on elaborating more fully a theory of how people seek information on politics, 
and how different electoral circumstances change this behaviour. 

\section{Acknowledgment}
We thank the Information Technology editors and reviewers for their very helpful comments. We also thank Wikimedia Deutchland e.V. and Wikimedia Foundation 
for the live access to the Wikipedia data via Toolserver and page view dumps.



\end{document}